\pdfoutput=1

\documentclass[11pt]{article}

\usepackage[]{ACL2023}

\usepackage{times}
\usepackage{latexsym}
\usepackage{hyperref}

\usepackage[T1]{fontenc}

\usepackage[utf8]{inputenc}

\usepackage{microtype}

\usepackage{inconsolata}

\usepackage{amsmath}
\usepackage{multirow}
\usepackage{booktabs}

\usepackage[normalem]{ulem}

\usepackage{amsmath,amsfonts,amsthm,bm} 

\usepackage{bm}

\newcommand{\mathtextbf}[1]{\text{\textbf{#1}}}

\usepackage[ruled, noline, fillcomment, ]{algorithm2e}
\usepackage{setspace}
\usepackage{etoolbox}

\usepackage{ amssymb }

\SetCommentSty{mycommfont}
\SetAlgoHangIndent{0.1em}

\SetKwComment{tcp}{$\cdot$\ }{}
\SetKwComment{tcc}{$\cdot$\ }{}

\SetAlCapNameSty{mycapsty}
\makeatletter
\renewcommand{\fnum@algocf}{\AlCapSty{\AlCapFnt}}
\makeatother

\SetAlgoCaptionSeparator{}



\makeatletter
\newcommand{\shortalgrule}{\par \vskip .05\baselineskip \moveright 0.07\columnwidth \vbox{\hrule width0.86\columnwidth}
\par \vskip .05\baselineskip}
\makeatother

\SetKwInOut{tempIn}{\mathtextbf{Template Inputs}}
\SetKwInOut{tempOut}{\mathtextbf{Template Output}}

\SetKwInOut{specific}{\mathtextbf{Specification}}
\SetKwInOut{condition}{\mathtextbf{Conditions}}

\makeatletter
\patchcmd{\@algocf@start}
  {-1.5em}
  {-0.4em}
  {}{}
\makeatother

\usepackage[most]{tcolorbox}
\usepackage{bm}
\usepackage{xfrac}
\usepackage{nicefrac}

\makeatletter
\newcommand{\eqnum}{\hfill\refstepcounter{equation}\textup{\tagform@{\theequation}}}

\makeatother

\newtcolorbox{eqbox}[1]{colback=black!1!white,colframe=black!70!white,fonttitle=\bfseries,
boxsep=2pt,left=5pt,right=3pt,top=2pt,bottom=2pt,title={\centering #1}}

\newtcolorbox{eqboxalgstyle}[1]{
colbacktitle=white,
colback=white,
colframe=black,
boxsep=2pt,left=5pt,right=3pt,top=2pt,bottom=2pt,
rightrule=-0.1pt, leftrule=-0.1pt,
arc=0pt,outer arc=0pt,
segmentation style={black!20},
title={\centering #1}}

\makeatletter
\everymath{\if@display\else\thickmuskip=2mu plus 2mu\fi}
\makeatother

\makeatletter
\newcommand{\eqboxlineshort}{\par \vskip .05\baselineskip \moveright 0.07\columnwidth \vbox{\hrule width0.86\columnwidth}
\par \vskip .05\baselineskip}
\makeatother

\usepackage{calc}
\usepackage{rotating}
\newcommand\mybox[1]{\centering \rotatebox{90}{%
    \parbox{\widthof{Performance Impact Per Noise\ }}{\centering #1}}}
    
\newcommand\myboxsm[1]{\centering \rotatebox{90}{%
    \parbox{\widthof{Reverb\ }}{\centering #1}}}
    
\newcommand\myboxmed[1]{\centering \rotatebox{90}{%
    \parbox{\widthof{Perception\ }}{\centering #1}}}

\title{Best Practices for Noise-Based Augmentation to Improve the Performance of Deployable Speech-Based Emotion Recognition Systems}

\author{Mimansa Jaiswal \\
  University of Michigan\\
  \texttt{mimansa@umich.edu} \\\And
  Emily Mower Provost \\
  University of Michigan\\
  \texttt{emilykmp@umich.edu} \\}

\begin{document}
\maketitle
\begin{abstract}
Speech emotion recognition is an important component of any human centered system. But speech characteristics produced and perceived by a person can be influenced by a multitude of reasons, both desirable such as emotion, and undesirable such as noise. To train robust emotion recognition models, we need a large, yet realistic data distribution, but emotion datasets are often small and hence are augmented with noise. Often noise augmentation makes one important assumption, that the prediction label should remain the same in presence or absence of noise, which is true for automatic speech recognition  but not necessarily true for perception based tasks. In this paper we make three novel contributions. We validate through crowdsourcing that the presence of noise does change the annotation label and hence may alter the original ground truth label. We then show how disregarding this knowledge and assuming consistency in ground truth labels propagates to downstream evaluation of ML models, both for performance evaluation and robustness testing. We end the paper with a set of recommendations for noise augmentations in speech emotion recognition datasets.
\end{abstract}

\section{Introduction}\label{sec:introduction}

%
%
%
%
Speech emotion recognition is increasingly included as a component in many real-world human-centered machine learning models. 
Modulations in speech can be produced for a multitude of reasons, both desirable and undesirable. In our case desirable modulations encode information that we want our model to learn and be informed by, such as  speaker characteristics or emotion. Undesirable modulations encode information that are extrinsic factors change with the environment, such as noise. In order to handle these modulations, we need large datasets that capture the range of possible speech variations and their relationship to emotion expression.  But, such datasets are generally not available for emotion tasks. To bridge this gap, researchers have proposed various methods to generate larger datasets.  One of the most common is noise augmentation. The baseline assumption of noise augmentation is that the labels of the emotion examples do not change once noise has been added~\cite{pappagari2021copypaste}. While this assumption can be confidently made for tasks such as automatic speech recognition (ASR), the same cannot be said for perception-based tasks, such as emotion recognition.

In this paper, we question the assumption that the annotation label remains the same in the presence of noise.  We first create a noise augmented dataset and conduct a perception study to label the emotion of these augmented samples, focused on the type of noise in samples whose perception has changed or remained the same given the agumentation. 
We use the results from this study to classify the complete set of augmentation noises into two categories, \textit{perception-altering} (i.e., noises that may change the perception of emotion) and  \textit{perception-retaining} (i.e., noises that do not change the perception of emotion). 
We propose that the perception-altering noises should not be used in supervised learning or evaluation frameworks because we cannot confidently maintain that the original annotation holds for a given sample. 
We evaluate the effects of disregarding emotion perception changes by examining how the performance of emotion recognition models and analyses of their robustness change in unpredictable manners when we include samples that alter human perception in the training of these models. Lastly, we provide a set of recommendations for noise based augmentation of speech emotion recognition datasets based on our results.

Researchers have considered the impact of noise on emotion perception and thereby the annotation of emotions. [X] looked at how pink and white noises in varying intensities change the perception of emotion. Another set of research has concentrated on training and validating noise robust models with the assumption that intent label prediction remains consistent in the presence of noise. For example, [X] have looked at training student teacher models that aim to ignore the effect of noise introduced to the model. On the other hand [X] have proposed copy pasting various emotion segments together along with neutral noise to balance the classes in an emotion dataset, thus improving performance.

In this paper, we claim that the standard assumption about perception and hence, label retention of emotion in the presence of noise may not hold true in a multiple noise categories. To understand which noises impact emotion perception, we use a common emotion dataset, IEMOCAP and introduce various kinds of noises to it, at varying signal to noise ratio (SNR) levels as well as at different positions in the sample. We then perform a crowdsourcing experiment that asks workers to annotate their perception of emotion for both the clean and the corresponding noise-augmented sample. This enables us to divide noise augmentation options into groups characterized by their potential to either influence or not influence human perception.

The results of the crowdsourcing experiments inform a series of empircal analyses focused on model performance and model robustness.  We first present an empirical evaluation of the effects of including perception-altering noises in training.  It will allow us to observe how the inclusion of perception-altering noises creates an impression of performance improvement.  We will discuss how this improvement is a myth, this new model will have learned to predict labels that are not truly associated with a given sample due to the perceptual effects of these noises.  
We consider both a general recurrent neural network (RNN) model and an end-to-end model for this purpose. We evaluate conditions in which novel augmentation noises are either introduced during training (matched) or seen for the first time during testing (mismatched).
The second empirical evaluation analyzes whether the gap in performance between the matched and mismatched conditions can be bridged using noise robust modeling techniques. 
The third and final evaluation is focused on the robustness of the model.  It will allow us to observe how the inclusion of these perception altering noises ultimately leads to a model that is more susceptible to attack compared to a model that does not include these noises.
We train an attack model for robustness testing. It considers a pool of noises and picks the best noise with a minimal SNR degradation that is able to change a model's prediction. We consider a condition in which the attack model has black-box access to the trained model.  The attack has a fixed number of allowed queries to the trained model, but not the internal gradients or structure (i.e., the attack model can only provide input and can only access the trained model's prediction). We test and  monitor the difference in the observed robustness of these aforementioned models.

We find that the crowdsourced labels do change in the presence of some kinds of noise. We then verify that the models perform worse on noisy samples when trained only on clean datasets. But, we show that this decrease in performance is different when using the complete set of noises for augmenting the test set vs. when only using the perception-retaining noises for augmentation. We show similar patterns for noise-robust models, specifically showing how there is an increased drop in performance for the end-to-end noise-robust model when excluding performance-altering noises during augmentation. We then discuss how our conventional metrics, those that look only at model performance, may be incorrectly asserting improvements as the model is learning to predict an emotion measure that is not in line with human perception. Troublingly, we find that the attack model is generally more effective when it has access to the set of all noises as compared to when excluding perception-altering noises for allowed augmentations. We also specifically find that given just a pool of carefully crafted reverberation modulations, the attack model can be successful in almost 65\% of the cases with minimal degradation in SNR and in less than ten queries to the trained model.
We end the paper with a general set of recommendations for noise augmentations in speech emotion recognition datasets.
\section{Research Questions}
In this paper, we investigate five research questions:

\textbf{Purpose 1: }Premise Validation through Crowdsourcing
\\
\textbf{RQ1: } Does the presence of noise affect emotion perception as evaluated by \emph{human raters}? Is this effect dependent on the utterance length, loudness, the type of the added noise, and the original emotion?
\\
\textbf{Reason: }Noise has been known to have masking effect on humans in specific situations. Hence, humans can often understand verbalized content even in presence of noise. Our goal is to understand whether the same masking effect extends to paralinguistic cues such as emotion, and to what extent. Our continuing claim from hereon remains that only noises that do not change human perception should be used for the training and evaluation of machine learning models. Not doing so, can lead to gains or drops in performance measurement that may not actually extend to real world settings. We call these changes "unverified" because we cannot, with certainity, be sure that the model should have predicted the original label (i.e., the label of the sample before noise was added) because the human did not neccessarily label the noisy instance with that same label.

\textbf{Purpose 2: }Noise Impact Quantification
\\
\textbf{RQ2: }  Can we verify previous findings that the presence of noise affects the performance of \emph{emotion recognition models}? Does this effect vary based on the type of the added noise? 
\\
\textbf{Reason: }We have known that presence of noise in data shifts the data distribution~\cite{chenchah2016speech}. This shift often leads to poor performance by machine learning models. We aim to quantify the amount of performance drop based on the type of noise in these systems, both, for any kind of noise, and then, specifically for noises that do not change human perception (perception-retaining).

\textbf{Purpose 3: }Denoising and Augmentation Benefits Evaluation
\\
\textbf{RQ3: } Does dataset augmentation (Q3a) and/or sample denoising (Q3b) help improve the robustness of emotion recognition models to unseen noise?
\\
\textbf{Reason: }We test whether the commonly-used methods for improving the performance of these models under distribution shifts is helpful. We focus on two main methods, augmentation and denoising. We specifically look at how performance changes when we augment with noises that include those that are perception-altering vs. when we exclude such noises.

\textbf{Purpose 4: }Model Robustness Testing Conditions
\\
\textbf{RQ4: } How does the robustness of a model to attacks compare when we are using test samples that with are augmented with perception-retaining noise vs. samples that are augmented with all types of noise, regardless of their effect on perception?
\\
\textbf{Reason: } Another major metric for any deployable machine learning algorithm is its performance on "unseen situations" or handling incoming data shifts (i.e., robustness testing). We test robustness using a noise augmentation algorithm that aims to forcefully and efficiently change a model's output by augmenting test samples with noise. We look at how often this algorithm is unsuccessful in being able to "fool" a model with its augmented samples. We look at the changes in frequency with which a model is successfully able to defend itself when the attack algorithm uses a set that includes all types of noises vs. when it only uses  perception-retaining noises.

\textbf{Purpose 5: }Recommendations
\\
\textbf{RQ5: } What are the recommended practices for speech emotion dataset augmentation and model deployment?
\\
\textbf{Reason: }We then provide a set of recommendations based on our empirical studies for deploying emotion recognition models in real world situations.

\section{Related Work}
\label{sec:prior}
Previous work that has focused on measuring how noise impacts the performance of machine learning models can be classified into three main directions: emotion recognition and noise-robust models, speech augmentation for classification purposes, and robustness testing in speech-based model training.
\subsection{Emotion Recognition}
End-to-End models are a recent paradigm for audio classification. The two major end-to-end models often used for speech classification tasks are: DeepSpeech~\cite{hannun2014deep} and wav2vec (1.0~\cite{schneider2019wav2vec}, 2.0~\cite{baevski2020wav2vec}).
Researchers have used the latest version of these pretrained models for recognizing speech emotion in various languages~\cite{pepino2021emotion, mohamed2021arabic}.
Researchers have also looked at extending traditional deep learning models to end-to-end models using multiple cojoined networks. For example,
Amirhossein et al. investigated how attention mechanisms could be processed at different layers of an end-to-end model to improve both speaker and emotion recognition~\cite{hajavi2021siamese}. 
In another approach, researchers have also analyzed how waveforms could be treated as concatenated image blocks and used this mechanism to perform real-time speech emotion recognition from an incoming audio stream~\cite{lech2020real}.
Other researchers have also looked at comparisons between performance of wav2vec2, wavBert, and HuBert, to understand which models perform the best for speech emotion recognition given a set of pre-known conditions~\cite{mohamed2021arabic}.

Researchers have explored techniques for noise-based data augmentation to improve noise robustness~\cite{kim2017acoustic}, focusing on how model training can be improved to yield better performance. However, these augmentation techniques tend to focus on acoustic event detection, speaker verification or speech recognition~\cite{ko2015audio}, and have been sparingly used in audio-based paralinguistic classification tasks. 

The common way to deal with noise in any audio signal is to use denoising algorithms. Hence, it is important to understand how machine learning models trained to recognize emotions perform if they are tested on denoised samples. Two common approaches include: Denoising Feature Space~\cite{valin2018hybrid} and Speech Enhancement~\cite{chakraborty2019front}. Denoising feature space algorithms seek to remove noise from the extracted front end features.  Speech enhancement algorithms seek to convert noisy speech to more intelligible speech. Both techniques are associated with challenges, from signal to harmonic dissonance~\cite{valin2018hybrid}.


\subsection{Noise Augmentation for Model Robustness}

\subsubsection{Noise Augmentation in Speech Recognition}
Research in speech recognition, rather than emotion recognition, has also tackled this problem.  Researchers have investigated methods to build speech recognition systems that are robust to various kinds and levels of noise~\cite{li2014an}. The common themes are a concentration on either data augmentation or gathering more real-world data to produce accurate transcripts~\cite{zheng2016improving}. Other lines of work have looked into preventing various attacks, e.g., spoofing or recording playback, on speaker verification systems~\cite{shim2018replay}. Noise in these systems is usually considered to be caused by reverberations or channel-based modulations~\cite{zhao2014robust}.  

\subsubsection{Noise Robustness in Emotion}
The general method in emotion recognition, like speech recognition, involves the addition of noise to the original sample. Researchers make use of publicly available and hierarchically categorized noise samples~\cite{chenchah2016speech} or introduce signal distortions to the sample itself. For example, researchers have looked at speed perturbation, additive white noise, vocal tract length perturbation and temp perturbation for augmenting speech emotion recognition datasets~\cite{nicolas2022data}.

The main concern with this approach is the underlying assumption that the addition of any noise in the background would not change the emotion label. It is important to note that most emotion recognition datasets are labelled as "perception of others" (i.e., annotated by an outside group of obervers, rather than the speaker themself) and are not self-reported. Intuitively, one might think that we should still be able to predict the ``correct'' label in the presence of noise because the speaker's emotion did not change. But given that the labels do not represent a speaker's evaluation of their emotion, rather, fall into the category of how others perceive another person's emotion, noise can (and does, as we show later and as supported by prior work in psychology~\cite{ma2015human}), change the emotion label.

Another recent approach, 
CopyPaste~\cite{pappagari2021copypaste}, takes a speaker-specific approach, augmenting samples from a single speaker with neutral examples from that speaker.  They argue that this approach will maintain the emotion label and, as a result of data augmentation, help improve performance. 

\subsection{Robustness Testing}
A separate line of research has used noise to evaluate the robustness of a given model. 
For example, researchers have focused on adversarial example generation, aiming to create audio samples that change the output of the classifier. However, these methods assume white-box access to the network and create modifications in either feature vectors or actual wav files
~\cite{carlini2018audio}. 
This generally results in samples that either have perceptible differences when played back to a human, or are imperceptible to a human but fail to attack the model when played over-air~\cite{carlini2018audio}. 

Model robustness to noise or an adversarial attack  can be evaluated by adding noise to the dataset and testing performance of the model. This method is commonly used for various tasks, such as, speech recognition, or speaker identification~\cite{abdullah2019practical}, whose perception is ideally independent of noise as discussed before.
The closest task to ours, where the ground truth varies based on noise introduction, is sentiment analysis. In this case, the adversarial robustness is usually tested by flipping words to their synonyms, such that the meaning of the text remains the same, and analyzing how the predictions of the model change~\cite{ebrahimi2017hotflip}. 
There is not a direct parallel for the acoustics of speech.
Hence, the introduction of noise for emotion recognition while assuring that the perception remains the same can be more difficult.

To the best of our knowledge, this is the first work that has studied the effect of different kinds of real-world noise and varying amounts of noise contamination on the human perception of emotion and the implication of training on these datasets from the perspective of machine performance and robustness to adversarial attacks.

\section{Datasets}
\label{sec:dataset}
For our study, we use the IEMOCAP dataset~\cite{busso2008iemocap}, created to explore the relationship between emotion, gestures, and speech. The data contain recordings from five pairs of actors (one male and one female), 10 actors in total.  The actors either performed from scripted scenes or improvised based on target scenarios. The data were segmented by speaker turn, resulting in a total of 10,039 utterances (5,255 scripted turns and 4,784 improvised turns). IEMOCAP contains audio, video, and associated manual transcriptions. 

The data were evaluated in terms of dimensional and categorical labels.  The dimensional evaluations included valence (positive vs. negative), activation (calm vs. excited), and dominance (passive vs. dominant).  Each utterance was evaluated by at least two evaluators.  The final dimensional labels are the average over the individual evaluators.  We bin the labels into three classes \textit{\{low, mid, high\}}, which we defined as \{(low:[1,2.75]), (mid:(2.75,3.25]), (high:(3.25,5])\}. 

\section{Noise}
\label {subsec:noise}

We investigate the effects of two types of noise, environmental and signal distortion. Environmental noises are additive, while signal distortion noise involves other types of signal manipulation.

\label{sec:noise}
\subsection{Environmental Noise}
We define environmental noises (\textit{ENV}) as additive background noise, obtained from the \href{https://github.com/karoldvl/ESC-50}{ESC-50 dataset}\cite{piczak2015dataset}.  ESC-50 is generally used for noise contamination and environmental sound classification ~\cite{xu2021head}. 
These environmental sounds are representative of many types of noise seen in real world deployments, especially in the context of virtual and smart home conversational agents. We use the following categories:
 
\begin{itemize}
    \item Natural soundscapes  (\textit{Nat}), e.g., rain, wind.
    \item Human, non-speech sounds (\textit{Hum}), e.g., sneezing, coughing, laughing or crying in the background etc.
    \item Interior/domestic sounds (\textit{Int}), e.g., door creaks, clock ticks etc.
\end{itemize}

We manipulate three factors when adding the noise sources: 
\begin{itemize}
    \item 
    \textit{Position}: The position of the introduction of sound that: (i) starts and then fades out in loudness or (ii) occurs during the entirety of the duration of the utterance. In the second case, this complete additive background would represent a consistent noise source in real world (e.g., fan rotation).
    \item \textit{Quality Degradation}: The decrease in the signal to noise ratio (SNR) caused by the addition of the additive background noise at levels of 20dB, 10dB and 0dB.  This is used only when noise is added to the entirety of the utterance.
\end{itemize}

\subsection{Signal Distortion}
We define signal distortion noise as modulations that aren't additive in the background. These kinds of noise in the audio signal can occur from linguistic/paralinguistic factors, room environment, internet lags, or the physical locomotion of the speaker. 

We use the nine following categories: 
\begin{itemize}
    \item \textit{SpeedUtt}: The utterance is sped up by either 1.25$\times$ or 0.75$\times$. 
    \item \textit{SpeedSeg}: A random segment within an utterance is sped up by 1.25$\times$.  The package \href{https://people.csail.mit.edu/hubert/pyaudio/}{pyAudio}
    that we used to speed up a segment did not permit slowing a segment down.  Thus, the 0.75$\times$ was not used here. 
    \item \textit{Fade}: The loudness of the utterance is faded by 2\% every second, which emulates the scenario of a user moving away from the speaker. The loudness is increased for fade in, and decreased for fade out. 
    \item \textit{Filler}: Non-verbal short fillers such as `uh', `umm' (from the same speaker) are inserted in the middle of a sentence. The insertion is either just the filler 
    or succeeded and preceded by a long pause 
 Fillers are obtained by parsing audio files for a given speaker and finding occurrences of any of the options from the above mentioned set. We will release the extracted fillers per speaker for IEMOCAP
    \item \textit{DropWord}: A randomly selected set of non-essential words belonging to the set: \{a, the, an, so, like, and\} are dropped from an utterance using word-aligned boundaries and stiching the audio segments together.
    \item \textit{DropLetters}: Following the same approach as drop word, letters are dropped in accordance with various linguistic styles chosen from the set: \{/\textbf{h}/+vowel, vowel+/n\textbf{d}/+consonant(next word), consonant+/\textbf{t}/+consonant(next word), vowel+/\textbf{r}/+consonant, /ihn\textbf{g}/\}. This is supported by research that has studied phonological deletion or dropping of letters in the native US-English dialect~\cite{phono, yuan2011automatic}. 
    \item \textit{Laugh/Cry}: ``Sob'' and ``short-laughter'' sounds are added to the utterance.  They are obtained from AudioSet~\cite{gemmeke2017audio}.  
    \item \textit{Pitch}: The pitch is changed by $\pm$ 3 half octaves using the pyAudio library. 
\item \textit{Rev}: Room reverberation is added to the utterance using py-audio-effects (\href{https://github.com/carlthome/python-audio-effects}{pysndfx}). We vary metrics such as reverberation ratio or room size to vary the type and intensity of reverberation added.
\end{itemize}

\subsection{Sampling and Noise-Perturbations}
\label{sec:datasample}
We randomly select 900 samples from the IEMOCAP dataset, which is far larger than the ones used for previous perception studies~\cite{parada-cabaleiro2017the, scharenborg2018the}. We select 100 samples from each activation and valence pair bin, i.e., 100 samples from the bin with activation: \emph{low}, valence: \emph{low}; 100 samples from the bin with activation: \emph{low}, and valence: \emph{mid}, and so on. This ensures that the chosen 900 samples cover the range of emotions expressed.  We impose another constraint on these 100 samples from each bin, 30 of them are shorter than the first quartile or greater than fourth quartile of utterance length in seconds to cover both extremities of the spectrum, and the remaining 70 belong in the middle. We also ensure that the selected samples had a 50-50 even split amongst gender.
We introduce noise to the 900 samples (Section~\ref{sec:dataset}).  Each sample is modulated in ten ways: four randomly chosen types of environmental noise and six randomly chosen signal distortion noise modulations, giving us a total of 9,000 noisy samples.

\section{User study}
\label{sec:userstudy}
We first analyze the effects of noise on human perception by relabeling the noise-enhanced data using the Amazon Mechanical Turk (AMT) platform.  We use insights from this experiment to guide the machine learning analyses that follow.

\subsection{Crowdsourcing Setup} 
We recruited 147 workers using Amazon Mechanical Turk who self-identify as being from the United States and as native English speakers, to reduce the impact of cultural variability.  We ensured that each worker had $>98$\% approval rating and more than 500 approved Human Intelligence Tasks (HITs). We ensured that all workers understood the meaning of activation and valence using a qualification task that asked workers to rank emotion content similar to~\cite{jaiswal2019muse}.  The qualification task has two parts: (i) we explain the difference between valence and activation and how to identify those, and, (ii) we ask them to identify which of the two samples has a higher/lower valence and a higher/lower activation, to ensure that they have understood the concept of activation and valence annotations. All HIT workers were paid a minimum wage ($\$9.45/$hr), pro-rated to the minute. Each HIT was annotated by three workers.

For our main task, we created pairs that contained one original and one modulated sample.  We then asked each worker to annotate whether or not they perceived the pair to have the same emotion. If they said \textit{yes} for both activation and valence, the noisy sample was labeled \textit{same} and they could directly move to the next HIT. If they said \textit{no}, the noisy sample was labeled \textit{different}. In this case, they were asked to assess the activation and valence of the noisy sample using Self Assessment Manikins \cite{bradley1994measuring} on a scale of [1, 5] (similar to the original IEMOCAP annotation). 

We also include three kinds of attention checks:
	\begin{enumerate}
		\item We show two samples that have not been modified and ask them to decide if the emotion represented was different. If the person says yes, then the experiment ends.
		\item We observe the time spent on the task. If the time spent on the task is less than the combined length of both samples, then the user's qualification to annotate the HITs is rescinded and their responses are discarded.
		\item We show two samples, one which has a gold standard label, and another, which has been contaminated with significant noise (performance degradation $>$30dB), such that the resulting sample is incomprehensible. If people do not mark this set of samples as being different, the experiment ends.
	\end{enumerate}
\noindent The failure rate based on the above  criteria was 8\%.
We ensured the quality of the annotations by paying bonuses based on time spent, not just number of HITs, and by disqualifying annotators
if they annotated any sample (including those outside of the attention checks) more quickly than the combined length of the audio samples.

We then created two sets of labels for each noise-augmented clip.  The \emph{first type} of label compared a noise-augmented clip to its original.  The noise-augmented clip was labeled the \emph{same} if the modified and original clip were perceived to have the same valence or activation, otherwise it was labeled \emph{different}.  We created this label by taking the majority vote over all evaluations.  The \emph{second type} of label included valence and activation.  A noise-augmented clip was given the average valence and activation over all evaluations.  

The inter-annotator agreement was measured using Cohen's kappa. Conventionally, when estimating Cohen's kappa, annotators are not considered as individuals, instead reducing annotators to the generic 1, 2, and 3. The challenge is that this often leads to artificially inflated inter-annotator agreement because individual characteristics and behavior of a particular worker are not taken under consideration~\cite{hoek2017evaluating}. We take a different approach, creating a table for the calculation of the statistic that considers annotators as individuals with separate entries for each clip, following the approach of~\cite{hoek2017evaluating}.  If an annotator didn't evaluate a given clip, the cell has a null (missing data) value.
We found that the Cohen's kappa was 79\% for activation and 76\% for valence.



\begin{algorithm}
\small
\caption{\scriptsize Pseudo-code for testing model robustness. \textit{Exit Code} is \textit{\textcolor{blue}{Success}} when the algorithm finds a noise-augmented version of the sample that the model changes prediction for. \textit{Exit Code} is \textit{\textcolor{red}{Failure}} when the model maintains its predictions over the any of the noise-augmented versions tried.\label{algo-attack}}
Randomly sample 1 noise variation from each category mentioned in Section~\ref{sec:noise}.\;
$numAttempts = 0$\;
\For{each noise in selected random noises:}
{Add noise to the sample such that the decrease in SNR is 1.\;
Get the classifier output with this new sample variation.\;
$numAttempts += 1$\;
\If{$numAttempts>k$}{\Return{Exit Code = \textcolor{red}{Failure}}}
\If {classifier output changes}{\Return{Exit Code = \textcolor{blue}{Success}}}
}
 
\For{each noise in selected random noises:}
{Add noise to the sample such that the decrease in SNR is 5.\;
Get the classifier output with this new sample variation.\;
$numAttempts += 1$\;
\If{$numAttempts>k$}{\Return{Exit Code = \textcolor{red}{Failure}}}
\If {classifier output changes}{\Return{Exit Code = \textcolor{blue}{Success}}}
\If {classifier output changes}
{\While{classifier output does not change}{
Iterate over all SNR decreases from 2-5\;
Get classifier output for the modified sample\;
$numAttempts += 1$\;
\If{$numAttempts>k$}{\Return{Exit Code = \textcolor{red}{Failure}}}
\If{classifier output changes}{\Return{Exit Code = \textcolor{blue}{Success}}}
}
}
}
 
\For{each noise in selected random noises:}
{Add noise to the sample such that the decrease in SNR is 10.\;
Get the classifier output with this new sample variation.\;
\If {classifier output changes}{\Return{Exit Code = \textcolor{blue}{Success}}}
\If {classifier output changes}
{\While{classifier output does not change}{
Iterate over all SNR decreases from 6-10
Get classifier output for the modified sample\;
$numAttempts += 1$\;
\If{$numAttempts>k$}{\Return{Exit Code = \textcolor{red}{Failure}}}
\If{classifier output changes}{\Return{Exit Code = \textcolor{blue}{Success}}}
}
}
}
 
\Return{Exit Code = \textcolor{red}{Failure}}

\end{algorithm}

\section{Methods}
\begin{table}
\small
\centering
\caption{Hyper-parameters used to select the best performing model on validation subset whilst training the traditional deep learning model.}
\label{tab:model_hyp1}
\begin{tabular}{ll}
\toprule
Hyper-parameter & Values\\
\midrule
\textbf {Traditional} & \\
\midrule
No. of Convolution Kernels & \{64, 128\} \\
Convolution Kernels Width & \{2\} \\
Number of Convolution Layers & \{5\} \\
Number of GRU layers & \{1, 2, 3\} \\
Pooling Kernel Width & \{2, 4\}\\
GRU Layers Width & \{32, 64\}\\
Number of Dense Layers & \{1, 2, 3\}\\
\midrule
\textbf{End to End} & \\
\midrule
No. of Dense Layers & \{1, 2\}\\
\bottomrule
\end{tabular}
\end{table}
We now describe the emotion recognition approaches, presenting two separate pipelines, one that relies upon direct feature extraction (Section~\ref{sec:dnn_features}) and the other that is end-to-end (Section~\ref{sec:dnn_endtoend}).  This allows us to investigate whether noise has a consistent effect.  We discuss approaches to improve noise robustness by training models with noise-augmented data or denoised data (Section~\ref{sec:noise_aug}).  Finally, we describe the setup and evaluation of the model robustness using an untargeted model misclassification test, 
which measures a model's fragility in terms of how likely it is that the model's decisions will change when specific types of noise are observed at test time (Section~\ref{sec:attack}).

\subsection{Creation of Data Partitions} 
We use a subject-independent five-fold cross validation scheme to select our train, test and validation sets.  In the first iteration, sessions 1-3 are used for training, session 4 is used as validation, and session 5 is used for testing.  This is repeated in a round-robin fashion, resulting in each session serving as a validation and a test fold. We also divide possible noises in two different categories based on results of crowdsourcing study (see Section~\ref{sec:rq1}).  The first category is \emph{perception-altering}, those that changed perception of humans and hence cannot be used for model training or evaluation with the old annotations.  The second category is \emph{perception-retraining}, those that did not change human perception, and hence, the model should produce no change in predictions when using those noise categories for sample augmentation. 

We use the noise categories (seeSection~\ref{subsec:noise}) in two varying circumstances.  The first category is \textit{matched}, where both the training and testing sets are augmented with same kinds of noise (e.g., both have nature-based sounds in them).  The second category is \textit{mismatched},
where the testing set is augmented with a noise \emph{category} not used for augmenting the training set (e.g., only the test set is augmented with nature-based noise while the train set is augmented with human or interior noises).

\subsection{Traditional Deep Learning Network}
\label{sec:dnn_features}

We first explore a common ``traditional'' deep learning network that is used in speech emotion recognition.  In this method we extract Mel Filterbank (MFB) features as input to a model composed of convolutional and gated recurrent unit (GRU) layers.

\subsubsection{Features}
We extract 40-dimensional Mel Filterbank (MFB) features using a 25-millisecond Hamming window with a step-size of 10-milliseconds using \href{https://github.com/jameslyons/python\_speech\_features}{python-speech-features}. Each utterance is represented as a sequence of 40-dimensional feature vectors. We $z$-normalize the acoustic features using parameters extracted from the training dataset.  During each cross-validation fold, the parameters are chosen from the training data and are applied to both the validation and testing data. 

\subsubsection{Network}
Our baseline network is a state-of-art single utterance emotion classification model which has been used in previous research~\cite{aldeneh2017pooling, khorram2017capturing, krishna2018study}. The extracted MFBs are processed using a set of convolution layers and GRUs (see Table~\ref{tab:model_hyp1} for the hyperparameters used for these layers). The output of these layers is then fed through a mean pooling layer to produce an acoustic representation which is then fed into a set of dense layers to classify activation or valence. 


\subsubsection{Training.} 
We implement the models using the Keras library~\cite{chollet2015}. We use a cross-entropy loss function for each task (e.g., valence or activation).  We learn the model parameters using the RMSProp optimizer. We train our networks for a maximum of 50 epochs and use early stopping if the validation loss does not improve after five consecutive epochs. Once the training process ends, we revert the network's weights to those that achieved the lowest validation loss. We repeat the experiment five times.  We report the results in terms of  Unweighted Average Recall (UAR, chance is 0.33), averaged over all test samples and five repetitions.  We compare the performance of different models or the same model in different noisy conditions/partitions using a paired t-test using the Bonferroni correction, asserting significance when $p\leq0.05$.

\subsection{End-to-End Deep Learning Networks}
\label{sec:dnn_endtoend}

Next, we explore a transformer-based model.  In this method the raw audio signal is used as input to a pre-trained and fine-tuned network and the emotion prediction is directly obtained as an output. These models do not require us to perform manual or domain knowledge-based extraction of features. They instead have a feature encoder component inside the model, which is dynamic in nature, and hence, can change its output for the same signal based on the dataset and nature of the task.

\subsubsection{Features}
For the end-to-end deep learning models, we do not need to extract audio features. Instead we rely on the network itself to both normalize and extract features, that are later passed onto the deeper layers of the network.
The feature set here is the original wav files that are not modified in any capacity. The eventual representations are of size 512, reproducing the setup in the state-of-the-art implementation~\cite{pepino2021emotion}.

\subsubsection{Network}
Our baseline network is the state-of-the-art wav2vec2.0 emotion recognition model~\cite{pepino2021emotion}. The wav2vec model is comprised of three parts: (i) a convolutional neural network (CNN) that acts as feature encoder, (ii) a quantizier module, and (iii) a transformer module.  The input to the model is raw audio data (16kHz) that is passed to a multi-block 1-d CNN to generate audio representations (25ms).  The quantizer is similar to a variational autoencoder that encodes and extracts features using a contrastive loss.  The transformer is used for masked sequence prediction and encodes the bi-directional temporal context of the features. 

We use the base model, which has not been fine-tuned for ASR (wav2vec2.0-PT). We then fine-tune the base model to predict the binned emotion labels.  We use the final representation of the output as an input to dense layers 
to produce the final output.

\subsubsection{Training}
We implement the model provided in the \href{https://speechbrain.readthedocs.io/en/latest/API/\\speechbrain.lobes.models.fairseq\_wav2vec.html}{speech brain library}. 
As in the other pipeline (Section~\ref{sec:dnn_features}), we use cross-entropy loss for each task and learn the dense layer parameters. Reproducing the state of the art model
~\cite{pepino2021emotion} = 2,021
We run the model for a maximum of eight epochs.  We revert the network's state to the one that achieved the lowest validation loss. We repeat this experiment five times.  Again, we use UAR and report the results averaged over both subjects and repetitions.

\subsection{Noise Augmentation Overview}
\label{sec:noise_aug}
We will be assessing a model's ability to classify emotion given either environmental or signal distortion noise.  
We perform two kinds of analysis, one when using the set of noises that includes those that do alter human perception, and another when only using noises that are perception-retaining. We report overall model performances for both of these categories. 

For a more thorough analysis, we then
specifically focus on the categories of noise that do not significantly affect human perception.  This allows us to evaluate a model's robustness, or its fragility, with respect to variations that wouldn't alter a human's perception of emotion.  This is important because the overwhelming majority of the noise-augmented utterances in the IEMOCAP dataset were not included in the user study and, therefore, do not have perceptual labels (Section~\ref{sec:userstudy}).  We consider three types of environmental noise \{Human (Hum), Interior (Int), Natural (Nat)\} and three types of signal distortion noise \{Speeding a segment (SpeedSeg), Fade,  Reverberation (Reverb)\}.

We use two separate testing paradigms: (i) matched testing, in which all noise types are introduced to the training, testing, and validation data and (ii) mismatched testing, in which $n$-1 types of noise are introduced to the training and validation sets and the heldout type of noise is introduced to the test set. In all cases, we analyze the test data in terms of specific noise categories.  Therefore, the test sets are the same between the two paradigms.

We run both the matched and mismatched experiments twice, first with the noise-augmented data and second with a noise-robust/denoising pipeline.  The first iteration will allow us to quantify the effect of the noise on the traditional and end-to-end classification pipelines.  We then repeat the experiment with either denoised data for the traditional classifier  (Section~\ref{sec:denoise}) or using the noise-robust implementation of wav2vec2.0 for the end-to-end classifier (Section~\ref{sec:noiserobust}).  This allows us to investigate how, or if, noise-robust implementations can offset the effects of background noise.

\subsubsection{Denoising}
\label{sec:denoise}
We implement denoising using the well-known Recurrent Neural Network Noise Suppression (RNNNoise, denoising feature space) approach, proposed in 2017 for noise suppression~\cite{valin2018hybrid}.  RNNNoise is trained on environmental noise, and these noises overlap considerably with those in our dataset. We use the algorithm's default parameters  
and use it on an `as-is' basis for our experiments. We assume that the system does not have the knowledge of which noise, from the set of available noise categories, is introduced and, therefore, we do not compare with other denoising algorithms that assume a priori knowledge of noise category. 
The result is a set of `noise-suppressed' samples in the training, validation and testing sets. 

We pass all the data, including both the original and noise-augmented data, through a denoising algorithm.  This allows us to ensure that acoustic artifacts, if any, are introduced to both the original and noise-augmented data.  We then train the traditional deep learning model as described in Section~\ref{sec:dnn_features}. 

\subsubsection{Using a Noise-Robust Model}
\label{sec:noiserobust}
In the end-to-end model, we need to use a different denoising approach because the approach described in the previous section does not return a wav file, but instead is applied to the feature-space directly. 
Here, we enforce robustness to noise using a model trained to be noise-robust in an end-to-end fashion. We use the noise-robust version (Wav2Vec2-Large-Robust) of the aforementioned wav2vec2.0 model~\cite{hsu2021robust}. 
The noise-robust large model was pretrained on 16kHz sampled speech audio. Noisy speech datasets from multiple domains were used to pretrain the model: Libri-Light, CommonVoice, Switchboard, and, Fisher~\cite{hsu2021robust}. 
We then train the end-to-end model as described in Section~\ref{sec:dnn_endtoend}. 


\subsection{Model Robustness Testing}
\label{sec:attack}
Deployed emotion recognition models must be robust. One of the major scenarios that we robustness test  any speech-based model for is the presence of noise. But the set of noises we choose to test robustness on can lead  to different conclusions about the robustness of the models. In our case, we consider two different scenarios:
\begin{enumerate}
    \item Robustness evaluation when using perception-retaining samples, noise samples that do not change human perception
    \item Robustness evaluation when using any kind of noise (i.e., both perception-retaining and perception-altering)
\end{enumerate}

We perform robustness evaluation of a model by using the model's output predictions to create new noise-enhanced samples that change the model's output, compared to the original clean sample.  We do this using an untargeted model misclassification test, in which we add noise to the samples. The intentional misclassification algorithm assumes black-box model access. For our purposes, it needs to have access to: (i) a subset of the dataset, (ii) noises to add to create perturbed samples, and (iii) model input and output.

As in any other perturbation-based robustness testing, the goal is to introduce perturbations to the samples such that the resulting samples are as close to the original sample as possible. 
The minimally perturbed sample should be the one that causes a classifier to change its original classification. We measure the amount of perturbation using SNR, calculated using the logarithmic value of the ratio between the original signal and the noise-augmented signal's power. We note that the lower the decrease in SNR, the more minimally perturbed a sample is. The maximal decrease in SNR that we use in the algorithm is a difference of 10 dB. This condition ensures that the sample is not audibly judged as contaminated by humans~\cite{kidd2016determining}.

The algorithm to choose this minimally perturbed sample has four major components:
\begin{enumerate}
	\item Requirements: some labelled samples, noise files, model input and output access, unlabelled samples for testing, and, optionally, correlation between noise type and performance degradation for a given model.
	\item Looping: The algorithm then loops over each noise category to figure out whether it can successfully force the model to misclassify. The noise category order is random if we do not have access to the optional performance degradation correlations.
	\item Minimizing: The algorithm then aims to find the lowest decrease in dB, such that the model still misclassifies. This ensures that the resultant noisy sample is as imperceptible to humans as possible.
	\item End Condition: The algorithm ends if a noise addition has been found, or if it runs out of number of tries allowed for model access.
\end{enumerate}

Please see Algorithm~\ref{algo-attack} for more details. In the algorithm, \textit{numAttempts} is the number of times the algorithm is allowed to access the model's input-output pairs. \textit{Classifier output} refers to  the prediction made by the model when the attack algorithm sends an input to the model to be classified. \textit{Classifier output changes} is true when the model predicted the emotion label differently after noise was added to the sample, compared to the original clean sample. \textit{\textcolor{blue}{Success}} implies that the algorithm was successfully able to force the model to misclassify a particular sample in the allowed number of attempts. \textit{\textcolor{red}{Failure}} implies that the algorithm could not force the model to misclassify in the allowed number of attempts and that the model can be considered robust for that sample.

We use the above algorithm in two different settings, under two different pre-known assumptions, with four levels of allowed queries, and four models (64 categories):
\begin{enumerate}
    \item Settings $\{$All vs. Not-Altering Human Perception$\}$
    \item Pre-Known Assumptions $\{$No Knowledge vs.  Knowledge About Noise Category Degradation Level$\}$
    \item Allowed Queries $\{$5, 15, 20, inf$\}$
    \item Models $\{$Traditional Deep Neural Network (T-DNN), End to End Deep Neural Network (E-DNN), Noise-Robust T-DNN (T-DNN-NR), Noise-Robust E-DNN (E-DNN-NR)$\}$
\end{enumerate}

For all the test samples, we execute five runs of the above algorithm to account for randomization in noise choices. These five runs are then averaged to obtain the average success of misclassification or average robustness for a given sample (1- average success of misclassification). We then average the robustness value over all the test samples. We report our obtained results for the above mentioned scenarios.

\begin{table}
\small
\centering
\caption{The table shows the ratio of samples marked by human evaluators as imperceptible to difference in emotion perception. \textit{V: Valence, A: Activation, $\delta$V:Average change in Valence on addition of noise, $\delta$A:Average change in Activation on addition of noise.}}
\label{tab:model_res}
\begin{tabular}{llllll}
\toprule
\multicolumn{2}{l}{}                    & V    & A  & $\delta$V & $\delta$A    \\ 
\midrule
\multicolumn{6}{l}{\textbf{Environmental Noise}}       \\ 
\midrule
NatSt      &                            & 0.01 & 0.00  \\
NatdB (Co) & -10dB                      & 0.01 & 0.00  \\
           & Same                       & 0.02 & 0.00  \\
           & +10dB                      & 0.03 & 0.01  \\ 
\hline
HumSt      &                            & 0.01 & 0.00  \\
HumdB (Co) & -10dB                      & 0.03 & 0.00  \\
           & Same                       & 0.02 & 0.00  \\
           & +10dB                      & 0.04 & 0.01  \\ 
\hline
IntSt      &                            & 0.05 & 0.01  \\
IntdB (Co) & -10dB                      & 0.02 & 0.01  \\
           & Same                       & 0.02 & 0.00  \\
           & +10dB                      & 0.04 & 0.01  \\ 
\midrule
\multicolumn{6}{l}{\textbf{Signal Distortion}}           \\ 
\midrule
SpeedSeg   &                            & 0.01 & 0.0   \\ 
\hline
Fade       & In                         & 0.04 & 0.01  \\
           & Out                        & 0.04 & 0.00  \\ 
\hline
DropWord      &                            & 0.01 & 0.00  \\ 
\hline
DropLetters     &                            & 0.01 & 0.00  \\ 
\hline
Reverb     &                            & 0.04 & 0.01  \\ 
\midrule
Filler     & L                          & 0.10 & 0.06  \\
           & S                          & 0.06 & 0.03  \\ 
\hline
Laugh      &                            & 0.16 & 0.17  & + .11 & + .26\\ 
\hline
Cry        &                            & 0.20 & 0.22  & - .20 & - .43\\ 
\hline
SpeedUtt   & 1.25x                      & 0.13 & 0.03 & - .10 & - .13 \\
           & 0.75x                      & 0.28 & 0.06 & - .18 & - .23 \\ 
\hline
Pitch      & 1.25x                      & 0.22 & 0.07 & - .11 & + .19 \\
           & 0.75x                      & 0.29 & 0.10 & - .07 & - .15 \\
\bottomrule
\end{tabular}
\end{table}

\begin{table*}
\small
	\caption{State of the art model performance in terms of UAR when using the general versions of traditional deep learning and end-to-end deep learning models. No noise refers to clean speech, all noises refers to the combined set of perception-retaining and perception-altering noise.  The environmental and signal distortion categories shown include only the perception-retaining noises.  As a reminder, samples in the all noises category have an uncertain ground truth, the row is marked with two stars ($\ast\ast$). \textit{V: Valence, A: Activation, Clean: Training on clean dataset, Clean$+$Noise | Mismatch: Cleaning on noisy dataset and testing on mismatched noisy partition, Clean$+$Noise | Match: Cleaning on noisy dataset and testing on matched noisy partition.} Random chance UAR is 0.33. 
    }
	\renewcommand{\arraystretch}{1.3}
	\setlength{\tabcolsep}{0.5em}
	\centering
	\label{tab:sota}
	\begin{tabular}{c|ccccccccccccccccccc} \toprule
	\multicolumn{1}{c}{} &  &  &  &  & \multicolumn{7}{c}{\textbf{Traditional Deep Neural Network}} &  & \multicolumn{7}{c}{\textbf{End-To-End Deep Neural Network}} \\ \cmidrule[\heavyrulewidth]{1-20}
	\multicolumn{1}{c}{} &  &  &  &  & \multicolumn{2}{c}{\multirow{2}{*}{\textbf{Clean}}} &  & \multicolumn{4}{c}{\textbf{Clean+Noise}} &  & \multicolumn{2}{c}{\multirow{2}{*}{\textbf{Clean}}} &  & \multicolumn{4}{c}{\textbf{Clean+Noise}} \\ \cline{9-12}\cline{17-20}
	\multicolumn{1}{c}{} &  &  &  &  & \multicolumn{2}{c}{} &  & \multicolumn{2}{c}{\textbf{\textit{Mismatch}}} & \multicolumn{2}{c}{\textbf{\textit{Match}}} &  & \multicolumn{2}{c}{} &  & \multicolumn{2}{c}{\textbf{\textit{Mismatch}}} & \multicolumn{2}{c}{\textbf{\textit{Match}}} \\ 
	\cmidrule(lr){6-7}\cmidrule(lr){9-10}\cmidrule(lr){11-12}\cline{14-15}\cmidrule(lr){17-18}\cmidrule(lr){19-20}
	\multicolumn{1}{c}{} &  &  &  &  & A & V &  & A & V & A & V &  & A & V &  & A & V & A & V \\ \midrule
	
	\multicolumn{4}{c|}{\textbf{No Noise}} &  &  
	
	0.67 &  0.59 &  & - & - & - & - &  & 
	
	0.70 & 0.63 &  & - &-  & - & - \\ \midrule

 \multicolumn{4}{c|}{\textbf{All Noises**}} &  &  
	
	0.40 &  0.38 &  & 0.55 & 0.42 & 0.66 & 0.59 &  & 
	
	0.70 & 0.63 &  & 0.44 & 0.38  & 0.67 & 0.60 \\ 

	\multicolumn{4}{c|}{\textbf{Perception Retaining Noises}} &  &  
	
	0.50 &  0.42 &  & 0.57 & 0.48 & 0.60 & 0.52 &  & 
	
	0.53 & 0.45 &  & 0.60 & 0.50  & 0.62 & 0.54 \\ \midrule
	
	\multirow{12}{*}{\rotatebox[origin=c]{90}{\textbf{ Environmental Category}}} & \textbf{Nature} & At Start &\multicolumn{1}{c|}{} &  & 
	 0.50 & 0.45 &  & 0.61 & 0.53 & 0.63 & 0.55 &  & 
	
	0.56 & 0.48 &  & 0.64 & 0.55 & 0.66 & 0.58 \\
	
	 &  & \multirow{3}{*}{\rotatebox[origin=c]{90}{Cont.}} & \multicolumn{1}{||c|}{-5dB} &  & 
	 0.45 & 0.39 &  & 0.55 & 0.44 & 0.59 & 0.50 &  & 
	 
	 0.49 & 0.42 &  & 0.58 & 0.47 & 0.62 & 0.53 \\
	 
	 &  &  & \multicolumn{1}{||c|}{-10dB} &  & 
	 
	 0.42 & 0.35 &  & 0.56 & 0.46 & 0.59 & 0.49 &  & 
	 
	 0.47 & 0.38 &  & 0.57 & 0.48 & 0.61 & 0.51 \\
	 
	 &  &  & \multicolumn{1}{||c|}{-20dB} &  & 
	 
	 0.40 & 0.35 &  & 0.51 & 0.44 & 0.55 & 0.47 &  & 
	 
	 0.47 & 0.39 &  & 0.52 & 0.46 & 0.56 & 0.50 \\
	   \cmidrule{2-20}
	
	 & \textbf{Interior} & At Start & \multicolumn{1}{c|}{}&  & 
	 
	 0.53 & 0.44 &  & 0.61 & 0.52 & 0.64 & 0.57 &  & 
	 
	 0.57 & 0.47 &  & 0.64 & 0.56 & 0.66 & 0.59 \\
	 
	 &  & \multirow{3}{*}{\rotatebox[origin=c]{90}{Cont}} & \multicolumn{1}{||c|}{-5dB} &  & 
	 0.46 & 0.36 &  & 0.55 & 0.44 & 0.58 & 0.49 & &
	 0.49 & 0.39 &  & 0.59 & 0.49 & 0.62 & 0.51 \\

	 &  &  & \multicolumn{1}{||c|}{-10dB} &  & 
	 
	 0.44 & 0.36 &  & 0.54 & 0.43 & 0.57 & 0.48 &  & 
	 
	 0.49 & 0.39 &  & 0.56 & 0.45 & 0.59 & 0.52 \\
	 
	 &  &  & \multicolumn{1}{||c|}{-20dB} &  & 
	 
	 0.40 & 0.35 &  & 0.52 & 0.44 & 0.55 & 0.49 &  & 
	 
	 0.46 & 0.37 &  & 0.56 & 0.45 & 0.58 & 0.51 \\
	   \cmidrule{2-20}
	
	 & \textbf{Human} & At Start & \multicolumn{1}{c|}{} &  & 
	 
	 0.52 & 0.45 &  & 0.60 & 0.51 & 0.63 & 0.55 &  & 
	 
	 0.58 & 0.47 &  & 0.62 & 0.52 & 0.66 & 0.57 \\
	 
	 &  & \multirow{3}{*}{\rotatebox[origin=c]{90}{Cont}} & \multicolumn{1}{||c|}{-5dB} &  & 
	 
	 0.45 & 0.37 &  & 0.52 & 0.43 & 0.55 & 0.48 &  & 
	 
	 0.49 & 0.40 &  & 0.56 & 0.44 & 0.57 & 0.50 \\
	 
	 &  &  & \multicolumn{1}{||c|}{-10dB} &  & 
	 
	 0.42 & 0.34 &  & 0.51 & 0.43 & 0.53 & 0.47 &  & 
	 
	 0.49 & 0.38 &  & 0.53 & 0.45 & 0.55 & 0.49 \\
	 
	 &  &  & \multicolumn{1}{||c|}{-20dB} &  & 
	 
	 0.40 & 0.34 &  & 0.50 & 0.41 & 0.53 & 0.46 &  & 
	 
	 0.46 & 0.38 &  & 0.54 & 0.43 & 0.56 & 0.48 
	 
	\rule{0pt}{3ex}    \\
	 \cmidrule{1-20}
	
	\multirow{6}{*}{\rotatebox[origin=c]{90}{\textbf{ Signal Distortion }}} & 
	
	\multicolumn{3}{l|}{Speed Segment} &  & 
	
	0.61 & 0.52 &  & 0.63 & 0.53 & 0.64 & 0.55 &  & 
	
	0.63 & 0.55 &  & 0.64 & 0.55 & 0.67 & 0.58 \\
	
	 & \multicolumn{2}{l}{Fade} & \multicolumn{1}{||c|}{In} &  & 
	 
	 0.62 & 0.53 &  & 0.65 & 0.55 & 0.67 & 0.58 &  & 
	 
	 0.64 & 0.55 &  & 0.67 & 0.58 & 0.68 & 0.59 \\
	 
	 &  & & \multicolumn{1}{||c|}{Out} &  & 
	 
	 0.61 & 0.51 &  & 0.62 & 0.54 & 0.64 & 0.57 &  & 
	 
	 0.63 & 0.54 &  & 0.64 & 0.56 & 0.66 & 0.59 \\
	 
	 & \multicolumn{3}{l|}{DropWord} &  & 
	 
	 0.64 & 0.56 &  & 0.65 & 0.56 & 0.67 & 0.59 &  & 
	 
	 0.65 & 0.58 &  & 0.67 & 0.58 & 0.69 & 0.61 \\
	 
	 & \multicolumn{3}{l|}{DropLetters} &  & 
	 
	 0.65 & 0.58 &  & 0.69 & 0.60 & 0.71 & 0.62 &  & 
	 
	 0.66 & 0.59 &  & 0.72 & 0.60 & 0.74 & 0.63 \\
	 
	 & \multicolumn{3}{l|}{Reverb} &  & 
	 
	 0.43 & 0.37 &  & 0.50 & 0.43 & 0.53 & 0.45 &  & 
	 
	 0.35 & 0.34 &  & 0.51 & 0.42 & 0.55 & 0.46 \rule{0pt}{2ex} \\ 
    
	
	 \bottomrule
	
	\end{tabular}
\end{table*}

 \begin{table*}
 \small
	\caption{Noise-Robust (NR) state of the art model performance in terms of UAR when using the noise-robust versions of traditional deep learning and end-to-end deep learning models. No noise refers to clean speech, all noises refers to the combined set of perception-retaining and perception-altering noise.  The environmental and signal distortion categories shown include only the perception-retaining noises.  As a reminder, samples in the all noises category have an uncertain ground truth, the row is marked with two stars ($\ast\ast$). \textit{V: Valence, A: Activation, Clean: Training on clean dataset, Clean$+$Noise | Mismatch: Cleaning on noisy dataset and testing on mismatched noisy partition, Clean$+$Noise | Match: Cleaning on noisy dataset and testing on matched noisy partition, NR: Noise Robust versions of the corresponding models} Random chance UAR is 0.33. 
    }
	\renewcommand{\arraystretch}{1.3}
	\setlength{\tabcolsep}{0.5em}
	\centering
	\label{tab:noise-robust}
	\begin{tabular}{c|ccccccccccccccccccc} \toprule
	\multicolumn{1}{c}{} &  &  &  &  & \multicolumn{7}{c}{\textbf{NR-Traditional Deep Neural Network}} &  & \multicolumn{7}{c}{\textbf{NR-End-To-End Deep Neural Network}} \\ \cmidrule[\heavyrulewidth]{1-20}
	\multicolumn{1}{c}{} &  &  &  &  & \multicolumn{2}{c}{\multirow{2}{*}{\textbf{Clean}}} &  & \multicolumn{4}{c}{\textbf{Clean+Noise}} &  & \multicolumn{2}{c}{\multirow{2}{*}{\textbf{Clean}}} &  & \multicolumn{4}{c}{\textbf{Clean+Noise}} \\ \cline{9-12}\cline{17-20}
	\multicolumn{1}{c}{} &  &  &  &  & \multicolumn{2}{c}{} &  & \multicolumn{2}{c}{\textbf{\textit{Mismatch}}} & \multicolumn{2}{c}{\textbf{\textit{Match}}} &  & \multicolumn{2}{c}{} &  & \multicolumn{2}{c}{\textbf{\textit{Mismatch}}} & \multicolumn{2}{c}{\textbf{\textit{Match}}} \\ 
	\cmidrule(lr){6-7}\cmidrule(lr){9-10}\cmidrule(lr){11-12}\cline{14-15}\cmidrule(lr){17-18}\cmidrule(lr){19-20}
	\multicolumn{1}{c}{} &  &  &  &  & A & V &  & A & V & A & V &  & A & V &  & A & V & A & V \\ \midrule
	
	\multicolumn{4}{c|}{\textbf{No Noise}} &  &  
	
	0.67 &  0.59 &  & - & - & - & - &  & 
	
	0.70 & 0.63 &  & - &-  & - & - \\ \midrule

 \multicolumn{4}{c|}{\textbf{All Noises**}} &  &  
	
	0.44 &  0.40 &  & 0.58 & 0.44 & 0.68 & 0.60 &  & 
	
	0.50 & 0.40 &  & 0.50 & 0.40  & 0.72 & 0.61 \\ \midrule

	\multicolumn{4}{c|}{\textbf{Perception Retaining Noises}} &  &  
	
	0.52 &  0.44 &  & 0.59 & 0.50 & 0.61 & 0.51 &  & 
	
	0.55 & 0.48 &  & 0.61 & 0.52  & 0.63 & 0.54 \\ \midrule
	
	\multirow{12}{*}{\rotatebox[origin=c]{90}{\textbf{ Environmental Category}}} & \textbf{Nature} & At Start &\multicolumn{1}{c|}{} &  & 
	 0.54 & 0.49 &  & 0.63 & 0.55 & 0.63 & 0.55 &  & 
	
	0.57 & 0.50 &  & 0.66 & 0.55 & 0.67 & 0.56 \\
	
	 &  & \multirow{3}{*}{\rotatebox[origin=c]{90}{Cont.}} & \multicolumn{1}{||c|}{-5dB} &  & 
	 0.50 & 0.42 &  & 0.58 & 0.50 & 0.60 & 0.52 &  & 
	 
	 0.49 & 0.42 &  & 0.61 & 0.51 & 0.63 & 0.53 \\
	 
	 &  &  & \multicolumn{1}{||c|}{-10dB} &  & 
	 
	 0.48 & 0.38 &  & 0.59 & 0.48 & 0.61 & 0.51 &  & 
	 
	 0.50 & 0.46 &  & 0.63 & 0.52 & 0.65 & 0.53 \\
	 
	 &  &  & \multicolumn{1}{||c|}{-20dB} &  & 
	 
	 0.44 & 0.38 &  & 0.55 & 0.49 & 0.59 & 0.51 &  & 
	 
	 0.50 & 0.43 &  & 0.53 & 0.46 & 0.58 & 0.47 \\
	   \cmidrule{2-20}
	
	 & \textbf{Interior} & At Start & \multicolumn{1}{c|}{}&  & 
	 
	 0.53 & 0.44 &  & 0.61 & 0.52 & 0.65 & 0.54 &  & 
	 
	 0.58 & 0.52 &  & 0.63 & 0.56 & 0.66 & 0.58 \\
	 
	 &  & \multirow{3}{*}{\rotatebox[origin=c]{90}{Cont}} & \multicolumn{1}{||c|}{-5dB} &  & 
	 
	 0.46 & 0.36 &  & 0.55 & 0.44 & 0.59 & 0.47 &  & 
	 
	 0.51 & 0.42 &  & 0.58 & 0.48 & 0.62 & 0.52 \\
	 
	 &  &  & \multicolumn{1}{||c|}{-10dB} &  & 
	 
	 0.44 & 0.36 &  & 0.54 & 0.43 & 0.57 & 0.43 &  & 
	 
	 0.49 & 0.44 &  & 0.57 & 0.48 & 0.61 & 0.50 \\
	 
	 &  &  & \multicolumn{1}{||c|}{-20dB} &  & 
	 
	 0.40 & 0.35 &  & 0.52 & 0.44 & 0.55 & 0.46 &  & 
	 
	 0.46 & 0.40 &  & 0.55 & 0.48 & 0.58 & 0.49 \\
	   \cmidrule{2-20}
	
	 & \textbf{Human} & At Start & \multicolumn{1}{c|}{} &  & 
	 
	 0.52 & 0.45 &  & 0.60 & 0.51 & 0.63 & 0.52 &  & 
	 
	 0.59 & 0.49 &  & 0.63 & 0.53 & 0.66 & 0.56 \\
	 
	 &  & \multirow{3}{*}{\rotatebox[origin=c]{90}{Cont}} & \multicolumn{1}{||c|}{-5dB} &  & 
	 
	 0.45 & 0.37 &  & 0.52 & 0.43 & 0.55 & 0.46 &  & 
	 
	 0.49 & 0.42 &  & 0.56 & 0.48 & 0.58 & 0.50 \\
	 
	 &  &  & \multicolumn{1}{||c|}{-10dB} &  & 
	 
	 0.42 & 0.34 &  & 0.51 & 0.43 & 0.54 & 0.44 &  & 
	 
	 0.47 & 0.38 &  & 0.54 & 0.49 & 0.55 & 0.45 \\
	 
	 &  &  & \multicolumn{1}{||c|}{-20dB} &  & 
	 
	 0.40 & 0.34 &  & 0.50 & 0.41 & 0.52 & 0.43 &  & 
	 
	 0.43 & 0.38 &  & 0.54 & 0.45 & 0.58 & 0.49 \\ 
	 \cmidrule{1-20}
	
	\multirow{6}{*}{\rotatebox[origin=c]{90}{\textbf{ Signal Distortion }}} & 
	
	\multicolumn{3}{l|}{Speed Segment} &  & 
	
	0.63 & 0.55 &  & 0.65 & 0.57 & 0.67 & 0.58 &  & 
	
	0.66 & 0.58 &  & 0.67 & 0.59 & 0.67 & 0.60 \\
	
	 & \multicolumn{2}{l}{Fade} & \multicolumn{1}{||c|}{In} &  & 
	 
	 0.64 & 0.55 &  & 0.66 & 0.57 & 0.68 & 0.59 &  & 
	 
	 0.65 & 0.58 &  & 0.67 & 0.59 & 0.69 & 0.60 \\
	 
	 &  & & \multicolumn{1}{||c|}{Out} &  & 
	 
	 0.64 & 0.56 &  & 0.65 & 0.57 & 0.67 & 0.58 &  & 
	 
	 0.66 & 0.57 &  & 0.68 & 0.59 & 0.69 & 0.63 \\
	 
	 & \multicolumn{3}{l|}{DropWord} &  & 
	 
	 0.67 & 0.60 &  & 0.66 & 0.58 & 0.66 & 0.59 &  & 
	 
	 0.68 & 0.60 &  & 0.69 & 0.60 & 0.69 & 0.60 \\
	 
	 & \multicolumn{3}{l|}{DropLetters} &  & 
	 
	 0.67 & 0.60 &  & 0.65 & 0.62 & 0.66 & 0.60 &  & 
	 
	 0.68 & 0.64 &  & 0.69 & 0.60 & 0.69 & 0.60 \\
	 
	 & \multicolumn{3}{l|}{Reverb} &  & 
	 
	 0.48 & 0.41 &  & 0.56 & 0.47 & 0.58 & 0.48 &  & 
	 
	 0.52 & 0.40 &  & 0.55 & 0.45 & 0.60 & 0.45 \\ 

  \bottomrule
	
	\end{tabular}
\end{table*}

\begin{table*}
\small
	\centering
	\renewcommand{\arraystretch}{1.3}
	\caption{Success of misclassification attempts on different models with varying number of allowed attempts (lower is better). As a reminder, samples in the all noises category have an uncertain ground truth, the row is marked with two stars ($\ast\ast$). Reverberation (reverb) is a perception-retaining noise that is also analyzed separately.  \textit{Trad: Traditional Deep Learning Model, E2E: End to End deep learning model, NR: Noise Robust version of the deep learning model.}}
	\begin{tabular}{@{}ccc|cccc|cccc} \toprule
	
	& \multirow{2}{*}{ \myboxsm{\textbf{Noise Set}} } & \multirow{2}{*}{\myboxsm{\textbf{No. of Attempts}}  } & 
	\multicolumn{4}{c}{\textbf{Activation}} & \multicolumn{4}{c}{\textbf{Valence}} 
	\\ \cmidrule{4-7} \cmidrule{8-11}
	 & 
	 & 
	 & \textbf{Trad} & \textbf{E2E} & \textbf{NR-Trad} & \textbf{NR-E2E}  & 
	 \textbf{Trad} & \textbf{E2E} & \textbf{NR-Trad} & \textbf{NR-E2E} \\ 
	  \cmidrule[\heavyrulewidth]{1-3}
	 \cmidrule[\heavyrulewidth]{4-7}  \cmidrule[\heavyrulewidth]{8-11}
	
	\multicolumn{1}{c|}{\multirow{10}{*}{
	
	\mybox{\textbf{Performance Impact Per Noise Is Unknown}}}} & 
	
	\multirow{4}{*}{
	
	\rotatebox[origin=c]{90}{All Noises**}} & 5 &
	
	0.29 & 0.22 & 0.15 & 0.10 &  
	
	0.11 & 0.15 & 0.13 & 0.05 \\
	
	 \multicolumn{1}{c|}{}&  & 15 &   
	 
	 0.31 & 0.33 & 0.31 & 0.32 &   
	 
	 0.23 & 0.24 & 0.22 & 0.22 \\

	 \multicolumn{1}{c|}{}&  & 25 &   
	 
	 0.40 & 0.28 & 0.40 & 0.33 &   
	 
	 0.22 & 0.19 & 0.21 & 0.20 \\
	 
	 \multicolumn{1}{c|}{}&  & inf &   
	 
	 0.43 & 0.41 & 0.44 & 0.35 &   
	 
	 0.25 & 0.20 & 0.26 & 0.18 \\ 
	 
	 \cmidrule{2-3}\cmidrule(r){4-11}
	 \multicolumn{1}{c|}{}& \multirow{4}{*}{
	 
	 \myboxmed{Perception-Retaining}} & 5 &   
	 
	 0.18 & 0.11 & 0.07 & 0.05 &   
	 
	 0.11 & 0.05 & 0.02 & 0.02 \\

	 \multicolumn{1}{c|}{}&  & 15 &  
	 
	 0.25 & 0.12 & 0.25 & 0.15 &   
	 
	 0.14 & 0.08 & 0.18 & 0.10 \\
	 
	 \multicolumn{1}{c|}{}&  & 25 &   
	 
	 0.32 & 0.24 & 0.32 & 0.20 &   
	 
	 0.13 & 0.10 & 0.11 & 0.09 \\
	 
	 \multicolumn{1}{c|}{}&  & inf &   
	 
	 0.40 & 0.26 & 0.36 & 0.23 &   
	 
	 0.19 & 0.14 & 0.17 & 0.14 \\ 
	 
  \cmidrule(r){4-11}
	 
	 \multicolumn{1}{c|}{}& \multirow{2}{*}{
	 
	 \myboxsm{Reverb}} & 5 &   
	 
	 0.33 & 0.15 & 0.22 & 0.18 &   
	 
	 0.20 & 0.14 & 0.22 & 0.09 \\
	 
	 \multicolumn{1}{c|}{}&  & 15 &   
	 
	 0.40 & 0.23 & 0.30 & 0.21 &   
	 
	 0.30 & 0.13 & 0.34 & 0.16  
	 
	  \rule{0pt}{3ex}  \\ \midrule
	\multicolumn{1}{c|}{\multirow{10}{*}{
	
	\mybox{\textbf{Performance Impact Per Noise Is Known}}}} 
	
	& \multirow{4}{*}{
	
	\rotatebox[origin=c]{90}{All Noises**}} & 5 &   
	
	0.33 & 0.28 & 0.24 & 0.22 &   
	
	0.15 & 0.14 & 0.12 & 0.12 \\
	
	 \multicolumn{1}{c|}{}&  & 15 &   
	 
	 0.38 & 0.38 & 0.32 & 0.33 &   
	 
	 0.20 & 0.21 & 0.18 & 0.16 \\
	 
	 \multicolumn{1}{c|}{}&  & 25 &   
	 
	 0.52 & 0.32 & 0.44 & 0.37 &   
	 
	 0.25 & 0.16 & 0.18 & 0.16 \\
	 
	 \multicolumn{1}{c|}{}&  & inf &   
	 
	 0.54 & 0.42 & 0.46 & 0.41 &   
	 
	 0.24 & 0.20 & 0.22 & 0.22 \\ 
	 
	 \cmidrule{2-3}
   \cmidrule(r){4-11}
	 
	 & \multicolumn{1}{|c}{\multirow{4}{*}{
	 
	 \myboxmed{Perception-Retaining}}} & 5 &   
	 
	 0.29 & 0.16 & 0.22 & 0.13 &   
	 
	 0.14 & 0.10 & 0.15 & 0.08 \\
	 
	 \multicolumn{1}{c|}{}&  & 15 &   
	 
	 0.32 & 0.32 & 0.32 & 0.28 &   
	 
	 0.14 & 0.15 & 0.16 & 0.12 \\
	 
	 \multicolumn{1}{c|}{}&  & 25 &   
	 
	 0.47 & 0.30 & 0.47 & 0.29 &   
	 
	 0.22 & 0.18 & 0.23 & 0.16 \\
	 
	 \multicolumn{1}{c|}{}&  & inf &   
	 
	 0.51 & 0.36 & 0.50 & 0.32 &   
	 
	 0.22 & 0.19 & 0.25 & 0.17 \\ 
	 
  \cmidrule(r){4-11}
	 
	 \multicolumn{1}{c|}{}& \multirow{2}{*}{
	 
	 \myboxsm {Reverb}} & 5 &   
	 
	 0.38 & 0.22 & 0.33 & 0.22 &   
	 
	 0.28 & 0.20 & 0.31 & 0.21 \\
	 
	 \multicolumn{1}{c|}{}&  & 15 &   
	 
	 0.47 & 0.29 & 0.41 & 0.28 &   
	 
	 0.30 & 0.23 & 0.35 & 0.26 
	 \rule{0pt}{3ex}    \\
	 \bottomrule
	\end{tabular}
 \label{tab:robustnessperf}
\end{table*}

\section{Analysis}
\subsection{Research Question 1 (Q1): Does the presence of noise affect emotion perception as evaluated by \emph{human raters}? Is this effect dependent on the utterance length, loudness, the type of the added noise, and the original emotion?}
\label{sec:rq1}
We find that the presence of environmental noise, even when loud, rarely affects annotator perception, suggesting that annotators are able to psycho-acoustically mask the background noise in various cases, as also shown in prior work (e.g.,~\cite{stenback2016speech}).


We find that the addition of signal distortion noise alters human perception.  The reported change in valence and activation values is on a scale of -1 to 1 (normalized). The addition of laughter changes the activation perception of 16\% of the utterances, with an average change of +22\% (+.26). The valence perception is altered in 17\% of the utterances, with an average change of +14\% (+.11).  Similarly for crying, valence is altered in 20\% of the cases, with an average change of -21\% (-.20). Crying changes activation perception in 22\% of the cases, with an average change of -32\% (-.43).
Raises in pitch also alter the perception of emotion.  In 22\% of utterances, the perception of activation is changed.  This contrasts with the perception of valence, which was altered only in 7\% of utterances. In this scenario, activation increases by an average of 26\% (+.19), and valence decreases by 12\% (-.11). On the other hand, decreases in pitch change the perception of activation in 10\% of the cases and of valence in 29\% of the cases. In this scenario, activation decreases by an average of 16\% (-.15), and valence decreases by 7\% (-.07). This ties into previous work~\cite{busso2009shrikanth}, which looked into how changes and fluctuations in pitch levels influenced the perception of emotions.  Changes in the speed of an utterance affect human perception of valence in 13\% (average of -.13) of the cases when speed is increased, and 28\% (average of -.23) when speed is decreased. On the other hand, changes in the speed of an utterance do not affect activation as often, specifically, 3\% in case of increase and 6\% in case of decrease.

We ensured that our crowsdourcing samples had an even distribution over gender of the speaker and the length of the sample (see Section~\ref{sec:datasample}). We performed paired t-test to evaluate whether these variables influenced the outcome of emotion perception change in presence of noise. We found that the changes in perception were not tied to characteristics of the speakers. For example, there was no correlation between changes in perception and variables such as, the original emotion of the utterance, the gender of the speaker, and the length of the utterance.





The human perception study provides insight into how emotion perception changes given noise.  This also provides information about the potential effects of noise addition on model behavior.  In the sections that follow, we will evaluate how machine perception changes given these sources of noise. 

\subsection{RQ2: Can we verify previous findings that the presence of noise affects the performance of \emph{emotion recognition models}? Does this effect vary based on the type of the added noise?}  
We first assess the performance of the model on the original IEMOCAP data and  find that the traditional model obtains a performance of 0.67 UAR on the activation and 0.59 UAR on the valence task. On the other hand, the end-to-end model obtains a performance of 0.73 on activation and 0.64 on the valence task. We hypothesize that the wav2vec2 model has an added advantage of being trained to recognize word structures that can incorporate some paralinguistic/langauge information in the fine-tuned model.

Next, we augment the test samples of each fold with each of the noise types (Section~\ref{sec:noise}) and investigate how the performance of the model changes.
We include two cases: (i) only perception-retaining noises and,(ii) all noises.

In the first scenario, we do not include noise types that were found to  affect human perception (e.g., \textit{Pitch, SpeedUtt, Laugh}) because once these noises are added, the ground truth is no longer reliable.  This lack of reliable ground-truth data hinders the evaluation of the model's performance on these samples because the majority of the utterances were not part of the original crowdsourcing experiment and are thus unlabeled.  The remainder of this section focuses on the second scenario only.

We find that for matched train and test noise conditions, the traditional machine learning model's performance decreases by an average of 28\% for environmental noise while it drops by 32\% for signal manipulation. On the other hand, for end-to-end deep learning model, the model's performance decreases by an average of 22\% and 26\% for environmental and manipulated noises, respectively.  In mismatched noise conditions, the models' performance decreases by an average of 33\% for environmental noise, fading, and reverberation. There is also a smaller drop in performance for speeding up parts of the utterance and dropping words, showing the brittleness of these models. Table~\ref{tab:sota} reports the percentage change in performance when testing on noisy test data, compared to clean test data.

We see that the end-to-end deep learning model is less affected by environmental noise, but has a larger drop due to fading and reverberation. We observe a larger drop on performance when dropping words, which possibly can be attributed to the change in audio-structure and non-controllable feature extraction for this model.

In the second scenario, we observe a large drop in performance for both the traditional and the end to end machine learning model. For example, in the case of a traditional deep learning model, the valence prediction performance drops to a near-chance performance when including all kinds of noises (see Table~\ref{tab:sota}).

We specifically want to point out how the inclusion of all noises in the test conditions changes the observed model performance. Primarily, the models on an average seem to do 20\% worse than they would if we only consider noises that do not alter human perception. We note the discrepancy between the results of the two noise addition scenarios and that results should be described with respect to the perceptual effects of noise, if noise augmentation is used.

\subsection{RQ3a: Does dataset augmentation help improve the robustness of emotion recognition models to unseen noise?}
\label{sec:modelaug}
We first report results for only perception-retaining noises.
When the training datasets are augmented with noise, we observe an average performance drop of 26\% and 10\% for matched noise conditions when using the traditional and the end-to-end deep learning model, respectively. For the mismatched noise conditions, we observe an average performance drop of 31\% and 16\% for the traditional and end-to-end deep learning models, respectively.

Both models see improved performance when the training dataset is augmented with continuous background noise in the matched noise setting.
We find that data augmentation improves performance on mismatched noisy test data over a baseline system trained only on the clean IEMOCAP data.  For example, the end-to-end model tested on environmental noise-augmented dataset (as compared to traditional deep learning model), reduces the performance drop to nearly zero.  This improvement is particularly pronounced (an increase of 22\% as compared to when trained on the clean partition) when the environmental noise is introduced at the start of the utterance (e.g., when the test set is introduced with nature-based noises at the start of the utterance and the train set is introduced with human and interior noises at the start of the utterance). We speculate that the network learns to assign different weights to the start and end of the utterance to account for the initial noise. 

However, we find that in both matched and mismatched conditions, it is hard to handle utterances contaminated with reverberation, a common use case, even when the training set is augmented with other types of noise. We find that this improvement in performance is even more reduced when using the wav2vec model, alluding to the model's fragility towards data integrity/structural changes. This can be because reverberation adds a continuous human speech signal in the background delayed by a small period of time. None of the other kind of noise types have speech in them, and hence augmentation doesn't aid the model to learn robustness to this kind of noise.


Finally, we investigate the differences in model performance when we use all types of noise vs. those that are perception-retaining.  Specifically, we focus on the perception-altering noises because samples augmented with noises in this category no longer have a known ground truth.  We inquire as to whether the use of samples that alter perception may lead to the appearance of model performance improvement (note: appearance because the samples now have uncertain ground truth).  
To maintain equivalence, we ensure that the training and validation dataset sizes are equal even when they are augmented with more noise conditions.
We observe that many cases of performance improvement occur when the noises include those that are perception-altering (see ``Al noises'' in Table~\ref{tab:sota}).
We observe a difference of 12\% to 25\% between the numbers that we obtain for the perception-retaining noises vs. when not distinguishing between the two noise categories. This supports our claim that the choice between types of noises used for data augmentation during model training and performance evaluation affects the empirical observations and should be carefully considered.
We hypothesize that this improvement in performance may be due to the inherent nature of noises that change emotion perception, if they are perceptible enough to change emotion perception, then they may stand out enough that the model can adequately learn to separate them out and improve its prediction towards the original ground truth annotation.  However, if the noise alteration truly does change perception, then the model is learning to ignore this natural human perceptual phenomenon.  This may have negative consequences during model deployment.


\subsection{RQ3b: Does sample denoising help improve the robustness of emotion recognition models to unseen noise?}
In the \emph{matched training testing condition}, we find that the traditional deep learning model has an average performance of 0.57 across all the datasets and testing setups, while the end-to-end models do substantially better at 0.61 UAR. See Table~\ref{tab:noise-robust} for details.

In the \emph{mismatched training testing condition}, we find that for the traditional deep learning model, adding a denoising component to the models leads to a significant improvement when the original SNR is high (e.g., after continuous noise introduction the SNR decreases only by 10dB).  In this case, we see an average improvement of $23\%\pm3\%$ across all environmental noise categories, compared to when there is no denoising or augmentation performed. 

However, when the SNR decreases by 20dB, we observe a decline in performance when using the noise suppression algorithms.  We believe that this decline in performance is reflective of the mismatch in goals: the goal of noise suppression is to maintain, or improve, the comprehensibility of the speech itself, not necessarily highlight emotion.  As a result, it may end up masking emotional information, as discussed in~\cite{ma2015human}.  

We further show that the addition of a denoising component does not significantly improve performance in the presence of signal distortion noise (other than reverberation) as compared to the presence of environmental noise (noise addition).  For example, when samples were faded in or out or segments were sped up, the performance is significantly lower ($-28\%$) than when tested on a clean test set. However, we did see an improvement in the performance  for \emph{unseen} reverberation contaminated samples as compared to data augmentation (an average of $+36\%$). Finally, we observe a general trend of increase in emotion recognition performance for the combined dataset (noisy and non-noisy samples), as compared to when the model is trained on the clean training set, which supports the findings from previous dataset augmentation research~\cite{aldeneh2017using}.

For the end-to-end deep learning model, we use the noise-robust version. We find that the model is effective at countering environmental noise when trained on a dataset augmented with environmental noise, even in the mismatched condition.  The performance is equivalent to the model evaluated on the clean data.
We further delve into the amount of noise augmentation needed to achieve this equivalency.  We consider all of the original training data.  We augment a percentage of the training data, starting by augmenting a random sample of 10\% with perception-retaining noise and increasing by 10\% each time.  We find that we obtain equivalency after augmenting with only 30\% of the training data.
We compare this compares to the traditional model, in which all of the training data are noise-augmented and we still do not see equivalency. 

We separately consider the signal distortion noise samples.  These were not part of the training of the wav2vec2-Large-robust model.  However, this model only sees a 6\% loss in performance, where the traditional robust model saw a 
20\%
loss in performance.

However, as discussed in the original traditional model, the end-to-end noise-robust model also fails on reverberation-based contamination even when trained on a similarly augmented dataset (note that the denoised traditional model could effectively handle reverberation). We believe that this may be because the wav2vec model is trained on continuous speech and relies on the underlying linguistic structure of speech.  However, in reverberation, there is an implicit doubling of the underlying information, which is at odds with how this model was trained.  This may explain why it is not able to compensate for this type of signal manipulation.

Next, we analyze whether the perception category of noise used for data augmentation of the samples, in both, the train and test dataset influences the reported results for noise-robust model improvement. We find that there is a significant difference in performance when the testing dataset is augmented with any kind of noise vs. when augmented with perception-retaining noise. 
Specifically, we observe that the maximal gains in performance when testing on matched noisy conditions are for samples for which we do not know whether or not the ground truth holds (i.e., both noise categories). For example, when using the noise robust traditional deep learning model, where the test and train dataset is augmented with any type of noises, we observe a performance improvement, as compared to that using a clean train dataset, of 12\%. Similarly for noise robust end to end models, the performance improvement difference when using all noises vs. only perception retaining ones is 15\% for activation and 13\% for valence.  Again, this is a problem when we think about deploying models in the real world because although the perception of these emotion expressions may change, we are assuming that the system should think of these perception labels as rigid and unchanging.

\subsection{RQ4: How does the robustness of a model to attacks compare when we are using test samples that with are augmented with perception-retaining noise vs. samples that are augmented with all types of noise, regardless of their effect on perception?}

In this section, we aim to show the effects of noise augmentation in general and specifically highlight noise categories that do not alter human perception.  We will show that if we are not careful with the selection of our noise types, moving from noise that we know not to alter perception to noise that may, the resulting noise sources can not only impact the brittleness of models, but also lead to inaccurate evaluation metrics. We also specifically report robustness performance when using reverberation-based contamination, as we observed that it is the most likely noise category to degrade the robustness of the model.

We allow the decision boundary attack algorithm a maximum of five queries to create a noise augmented sample that will change the model output.  We find that if the attacker is also given perception-altering noises, compared to perception-retaining, it can more effectively corrupt the sample.  It achieves an increase in success rate from 35\% (only perception-retaining) to 48.5\% (all noise categories).
This increase in the success rate when perception-altering noises are included implies that the model does not remain robust when the effects of noise on human perception are not considered.

We next consider the type of noise (environmental vs. signal manipulation).  We find that the success rate of flipping a model's output is 18\% for noises belonging to the environmental category, which is generally a category of perception-retaining noise. The success rate of flipping a model's output is 37\% for all noises belonging to the signal manipulation category.
When we constrain our possible noise choices to perception-retaining signal manipulations, we see that the success rate of the intentional misclassification algorithm drops to 24\%.
On the other hand, we observe that when we also consider the signal manipulations that are perception-altering, the success rate of flipping a model output is 39\%.  See Table~\ref{tab:robustnessperf} for more details.

We previously discussed the fragility of end-to-end models towards reverberation-based noise contamination, noise that is perception-retaining for human evaluators.
Here, we specifically run an experiment to use only that particular noise category for the model fragility testing. 
If the attacker knows that the model is susceptible to reverberation-based prediction changes, the intentional misclassification algorithm can land on an optimal set of room and reverberation parameters in a maximum of five queries to be able to produce a flipped output for that particular sample. It achieves a success rate of 24\%, compared to 12\% for other perception-retaining noises.
The traditional noise-robust deep learning model is even more challenged, compared to the end-to-end model.  The number of queries required to flip the output is three, vs. five for the end-to-end model, suggesting that it is less robust.
This empirical evaluation is performed primarily to demonstrate how such noise inclusions can not only invalidate the ground truth but also lead to inaccurate and fragile benchmarking and evaluation of adversarial efficiency and robustness.

\subsection{RQ5: What are the recommended practices for speech emotion dataset augmentation and model deployment?}
We propose a set of recommendations, for both augmentation and deployment of emotion recognition models in the wild, that are grounded in human perception. For augmentation, we suggest that:

\begin{enumerate}
    \item  Environmental noise should be added to datasets to improve generalizability to varied noise conditions, whether using denoising, augmentation, or a combination of both.
    \item  It is good to augment datasets by fading the loudness of the segments, dropping letters or words, and speeding up small (no more than 25\% of the total sample length) segments of the complete sound samples in the dataset. But it is important to note that these augmented samples should not be passed through the denoising component as the denoised version loses emotion information.
    \item One should not change the speed of the entire utterance more than 5\% and should not add intentional pauses or any background noises that elicit emotion behavior, e.g., sobs or laughter.
\end{enumerate}
Regarding deployment, we suggest that:
\begin{enumerate}
    \item Noisy starts and ends of utterances can be handled by augmentation, hence, if the training set included these augmentations, there should be no issue for deployed emotion recognition systems.
    \item Reverberation is hard to handle for even augmented emotion recognition models. Hence, the samples must either be cleaned to remove the reverberation effect, or must be identified as low confidence for classification.
    \item Deploy complementary models that identify the presence of noise that would change a human's perception.
\end{enumerate}

\section{Conclusion}
In this work, we study how the presence of real world noise, environmental or signal distortion, affects human emotion perception. We identify noise sources that do not affect human perception, such that they can be confidently used for data augmentation. We look at the change in performance of the models that are trained on the original IEMOCAP dataset, but tested on noisy samples and if augmentation of the training set leads to an improvement in performance. We conclude that, unlike humans, machine learning models are extremely brittle to the introduction of many kinds of noise. While the performance of the machine learning model on noisy samples is aided from augmentation, the performance is still significantly lower when the noise in the train and test environments does not match. In this paper, we demonstrate fragility of the emotion recognition systems and valid methods to augment the datasets, which is a critical concern
in real world deployment.

\bibliography{main.bib}
\bibliographystyle{acl_natbib}

\end{document}